\documentclass[a4paper]{article}

\usepackage{se/LaTeX/INTERSPEECH2021}

\usepackage{anyfontsize}
\usepackage{t1enc}

\usepackage{graphicx}
\usepackage{amssymb,amsmath,bm}
\usepackage{textcomp}
\usepackage{color}
\usepackage{multirow}
\usepackage{url}
\usepackage[activate]{microtype}
\usepackage{subcaption}
\usepackage{soul}
\usepackage[activate]{microtype}
\usepackage{ulem}
\usepackage{hyperref}
\usepackage{cleveref}

\def\vec#1{\ensuremath{\bm{{#1}}}}

\PassOptionsToPackage{dvipsnames,table}{xcolor}
\usepackage{pgfplotstable}
\pgfplotsset{compat=1.14}

\title{Real-time Streaming Wave-U-Net with Temporal Convolutions for Multichannel Speech Enhancement}
\name{Vasiliy Kuzmin$^1$, Fyodor Kravchenko$^1$, Artem Sokolov$^{1,3}$, Jie Geng$^2$ }
\address{
  $^1$Huawei Russia Research Institute
  \\$^2$Huawei Nanjing Research Institute, China
  \\$^3$HSE University, Nizhniy Novgorod, Russia
  }
\email{kuzmin.vasiliy@huawei.com, kravchenko.fyodor@huawei.com, sokolov.artem@huawei.com, gengjie3@huawei.com}

\begin{document}

\maketitle

\begin{abstract}

\noindent In this paper we describe our work that we have done to participate in Task1 of ConferencingSpeech2021 challenge. This task set a goal to develop the solution for multi-channel speech enhancement in a real-time manner. We propose a novel system for streaming speech enhancement. We employ Wave-U-Net architecture with temporal convolutions in encoder and decoder. We incorporate self-attention in decoder to apply attention mask retrieved from skip-connection on features from down-blocks. We explore history cache mechanisms that work like hidden states in recurrent networks and implemented them in proposal solution. It helps us to run inference with chunks length 40 ms and Real-Time Factor ~0.4 with the same precision. 

\end{abstract}
\noindent\textbf{Index Terms}: Multi-channel speech enhancement, speech enhancement, wave-u-net, self-attention, deep learning

\section{Introduction}

Speech enhancement (SE) tries to separate out individual audio sources from an input mixture of clean audio and noise. Multi-channel approaches, generally, do it better than single-channel ones considering that they could process spatial information taken from time differences between signal reach separate items of microphone arrays (MA). Conventional approaches like multi-channel Wiener filter ~\cite{postfiltering2001} or beamforming ~\cite{beamforming2001} are progressively displaced by deep learning (DL) based techniques ~\cite{deepLearningSpeechEnhancement2018} or by mixture with them ~\cite{deepLearningAndMvd2016}. Deep learning approaches with U-Net ~\cite{unet2015} based architectures was very successfully on various computer vision tasks connected to image and video processing where we need to transform one image to another or extract topological information that we need. Furthermore, this architecture was transferred to speech related tasks close to SE as voice separation ~\cite{voiceSeparation2017} and speech enhancement itself ~\cite{speechDereverberation2018}. This architecture works well as with signal waveform as with its Short Time Fourier Transform (STFT) ~\cite{timeFrequencyDomainSE2018}. 

The version adopted for processing time domain signal called Wave-U-Net is also actively exploited nowadays for the same tasks ~\cite{waveUNetForSE2018}. To obtain real-time streaming speech enhancement the network should ignore the future frames and provide the Real Time Factor (RTF) < 1. Recurrent networks like unidirectional Long-Short Time Memory ~\cite{lstm1997} (LSTM) and Temporal Convolution Networks (TCN) could satisfy the first restrictions as they are based on casual convolutions which have zero look ahead ~\cite{tcnWaveUNet2019}. Moreover, convolutions are very fast and allow to build real-time systems. LSTM mechanisms provide hidden states retrieving and sequences could be handled step-by-step by passing previous states for new iteration. But for convolution this trick is not supported from-the-box. In ~\cite{transformerXL2019} the authors showed that history cache with hidden states could be saved and reused. Thus, it is also possible to fulfill inference with small pieces of stream, but it requires to "manually" fed on hidden states from previous iteration with the input. This paper proposes a neural network architecture based on Wave-U-Net with temporal convolutions (TC Wave-U-Net) which uses cache with hidden states for the inference. 

To make prediction more precise various self-attention mechanisms are incorporated in Wave-U-Net systems ~\cite{attentionForSE2019}. We also do it in the paper as casual convolution gives in to convolution with future context and extra elements are required to yield acceptable precise. 

This method was submitted to the ConferencingSpeech 2021 challenge\footnote{\href{https://tea-lab.qq.com/conferencingspeech-2021}{https://tea-lab.qq.com/conferencingspeech-2021}} in INTERSPEECH 2021 conference. Summarily, our primary contributions are the follows:
\begin{itemize}
    \item We developed the novel model for the multi-channel speech enhancement task that can be inferences real-time manner, have zero look ahead and based on successful U-Net architecture. This model have only 8.31 millions parameters and can be run on any device. 
    \item We applied historical context cache which allowed us to decrease receptive field during inference as well as decrease total number of float points operation. 
\end{itemize}
The rest of the paper is organised as follows. First, in the following section, we briefly review the background of the problem. In Section 3 we describe our Wave-U-Net based Temporal Convolution network with attentions. The details of of our experiments and training procedure are then presented in Section 4. We present the results and comparison with the baseline system in Section 5. Finally, in Section 6 we provide some conjectures as why and how to configure historical cache, further research in this direction followed by conclusion.

\section{Background}
During participating ConferencingSpeech 2021 challenge we aimed to solve Task1 which is Multi-channel speech enhancement with single microphone array. This task expects the noisy audio processing from single linear array with non-uniform distributed microphones. Real time factor considering less than 1.0 while running on device as well as participant should use frame of length 40 ms.  

A multi-channel speech enhancement problem could be described the following way:

Let take \(C\) - channel signal on the \(t\)  time step as \(Y_t = [y_t^{1},...,y_t^{c}]\)

We assume that \(y_t^{c}\) is given as the following mathematical expression:
\begin{equation}
  y_t^{c} =  \overline{x_t^{c}} + n_t^{c} = h^{c} \circledast x_t^{c} + n_t^{c}
  \label{eq2}
\end{equation}
where \(x_t^{c}\) the clean signal recorded by the \(C\)-th microphone, \(n_t^{c}\) is the additive noise signal and \(h^{c}\) is room impulse response (RIR). The convolution of RIR and dry signal \(x_t^{c}\) called reverberant signal \(\overline{x_t^{c}}\). The aim of Task1 is to estimate signal \(x_t^{orig}\), where  \(orig \in \{1,...,C\}\), by removing additive noise and by dereverberation for whole utterance  \(t = 1, ..., T\). The \(x_t^{orig}\) represents the original mono channel audio from multi-channel mixture \(Y_t\).
\section{Architecture: Wave-U-Net with temporal convolutions}

In this section, we explain the TC Wave-U-Net architecture with attentions and history cache we implemented for streaming inference.

\subsection{Structure of TC Wave-U-Net}

\begin{figure}[t!]
  \centering
  \includegraphics[width=\linewidth]{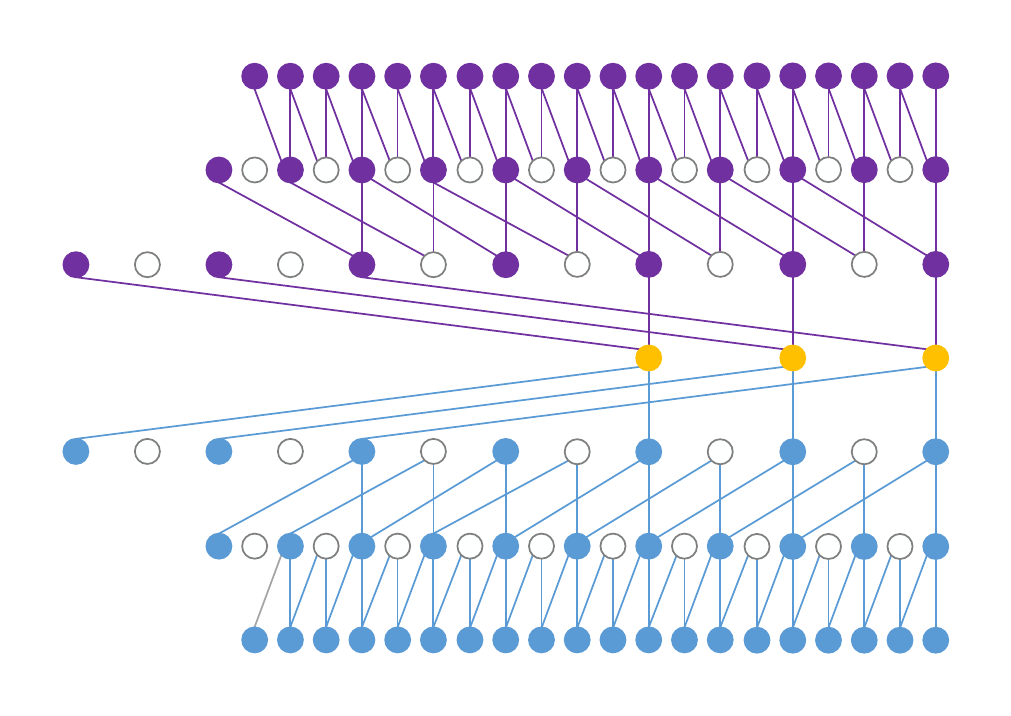}
  \caption{Schematic view of TC Wave-U-Net.}
  \label{fig:tc_unet}
\end{figure}

We follow the Wave-U-Net structure for our multi-channel speech enhancement solution. Our framework consists of an encoder, middle part (bottleneck) and decoder. We use history padding to avoid the non zero-look ahead. Schematically system is depicted on Figure~\ref{fig:tc_unet}. We note that our model takes C-channel waveform as its input and predicts an cleaned audio. 

\begin{figure}[t!]
  \centering
  \includegraphics[width=\linewidth]{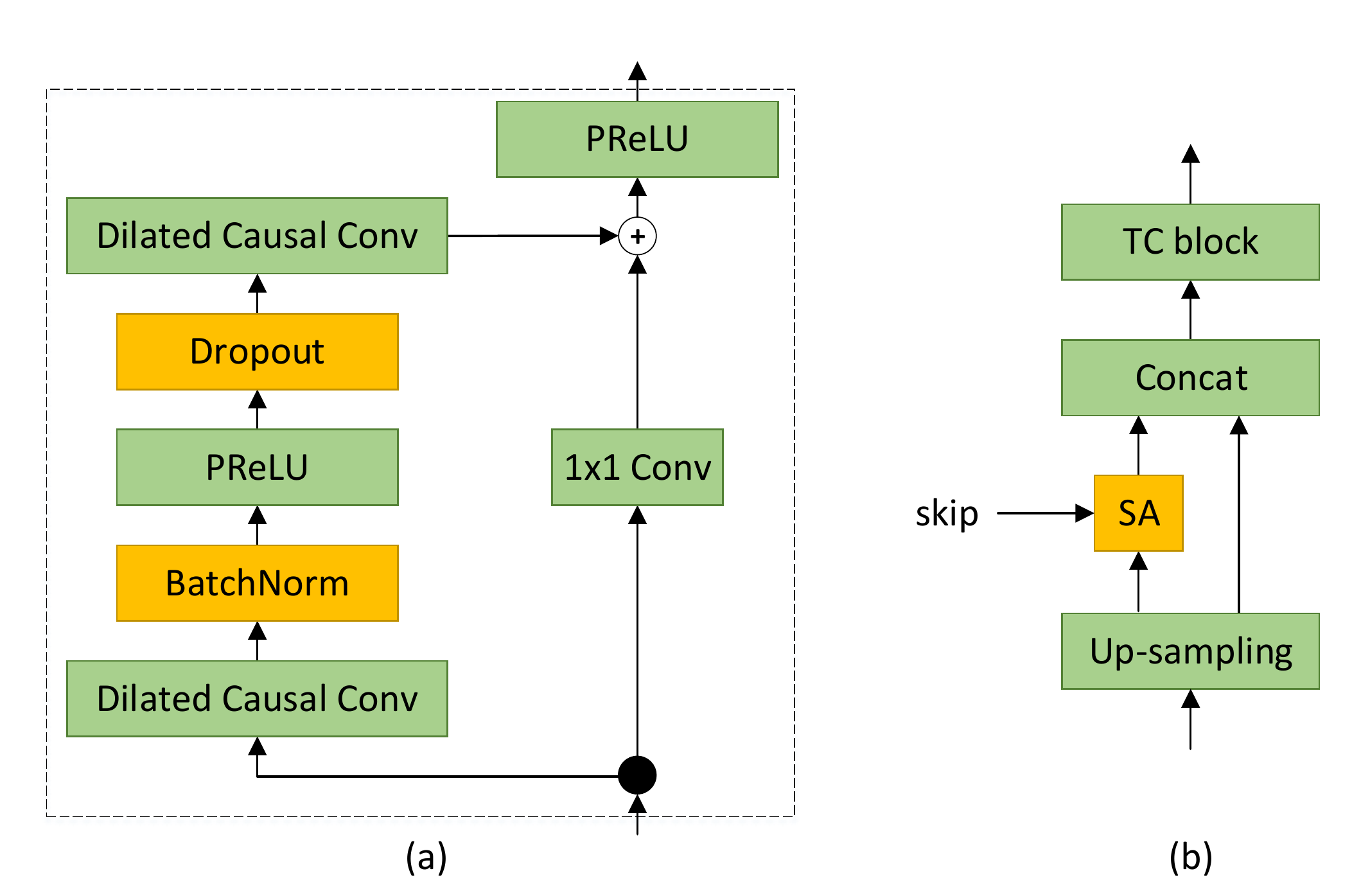}
  \caption{(a) TC Block, (b) Up Block. The skip connection is the
output from the corresponding down-block.}
  \label{fig:tc_and_down}
\end{figure}

Encoder and decoder consist of TC blocks. The block internals are visualised on Figure~\ref{fig:tc_and_down} (a). The block is a sequence of dilated casual 1d convolution, batch normalization, parametric non-linearity, dropout and one more convolution. A final Parametric Rectified Linear Unit ~\cite{prelu2015} (PReLU) is applied after additive operation of result with residual connection. We experimentally established that such architecture provides better convergence. We keep the SAME padding everywhere and after each TC block fulfill down-operation - reducing the time dimension on half just by removing of each second element. Our experiments showed that this kind of down sampling is more effective for our architecture than, for example, stride 2 in the convolution.
The decoder is organized symmetrically but it composed a bit more complicated as comprises attention mechanism (Figure~\ref{fig:tc_and_down}(b)). We use linear interpolation for \(2\times\) up-sampling. Skip connection between encoder and decoder blocks is organized as usual \(1D\) convolution.

The papers ~\cite{attentionForSE2019}, ~\cite{CAdenceUNetSE2020}, ~\cite{crossChannelWaveUNetSE2020} inspired us to integrate a self-attention to our architecture. We fed on self-attention an up block and skip connection tensors from the same hierarchical level. Afterwards, we concatenate Up Block output with self-attention output and result goes to TC block.

\begin{figure}[t!]
  \centering
  \includegraphics[width=\linewidth]{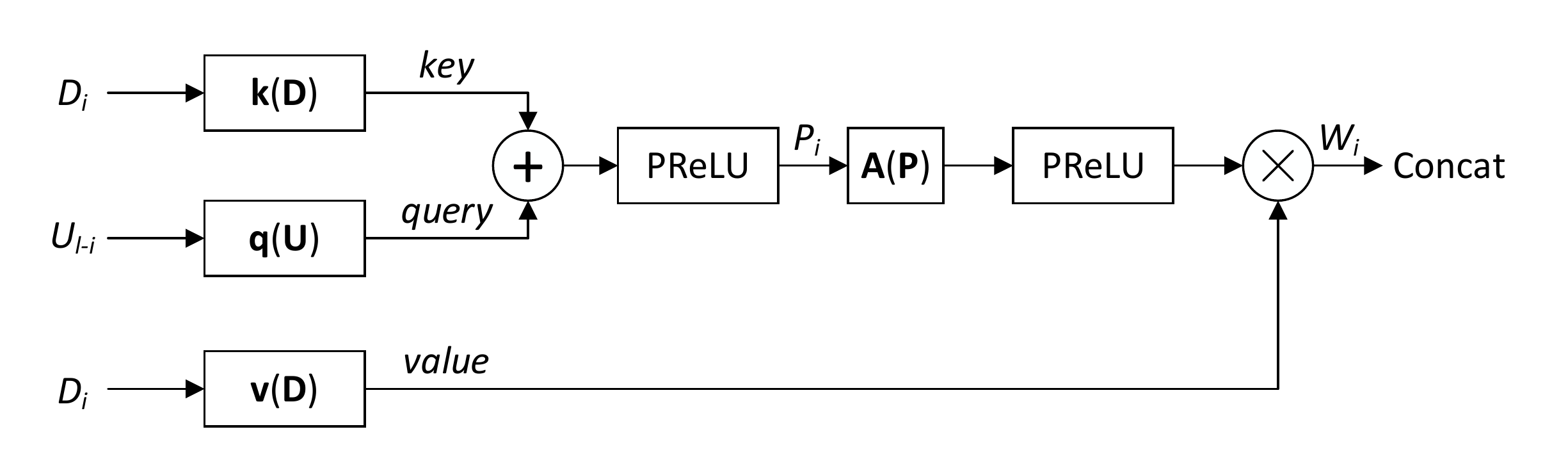}
  \caption{Proposed Self-Attention}
  \label{fig:attention}
\end{figure}

Self-attention mechanism is intended to capture the global dependencies. It is successfully exploit, for instance, in such fields of deep learning as speech recognition ~\cite{quartznet2020} and machine translation ~\cite{machineTranslationWithAttention2014}. In our architecture we employed attention by the similar way with ~\cite{attentionForSE2019} to find out relevant features came with skip connections by multiplying it with attention mask. The same approach is applied before the output of the network when original noisy signal and output from up-block are passed to attention. Visual explanation of proposed self-attention is shown Figure~\ref{fig:attention}. 

Let's give \(l\) the number of TC blocks in encoder/decoder. When \(D_i, i = 1,...l\) be output of down-blocks came as skip-connection and \(U_{l-i}\) be the output of linear interpolation. Attention uses \(\bf{k}\), \(\bf{q}\) and \(\bf{v}\) as \(1D\) convolution operations for the representation of input to embedding space. The products \(key\) and \(query\) are then summarised and, after parametric linearity, as product \(P_{i}\) follows to attention mask calculation \(A(P_{i})\) which is a convolution with kernel size 1. The output of attention is a term-wise product of \(W_i = A(P_{i})D_i\). In our experiments, attention block helps the network to converge deeper, while without it, a model converges faster, however it gets stuck in some space.

\subsection{Streaming Inference}
\begin{figure}[t]
  \centering
  \includegraphics[width=\linewidth]{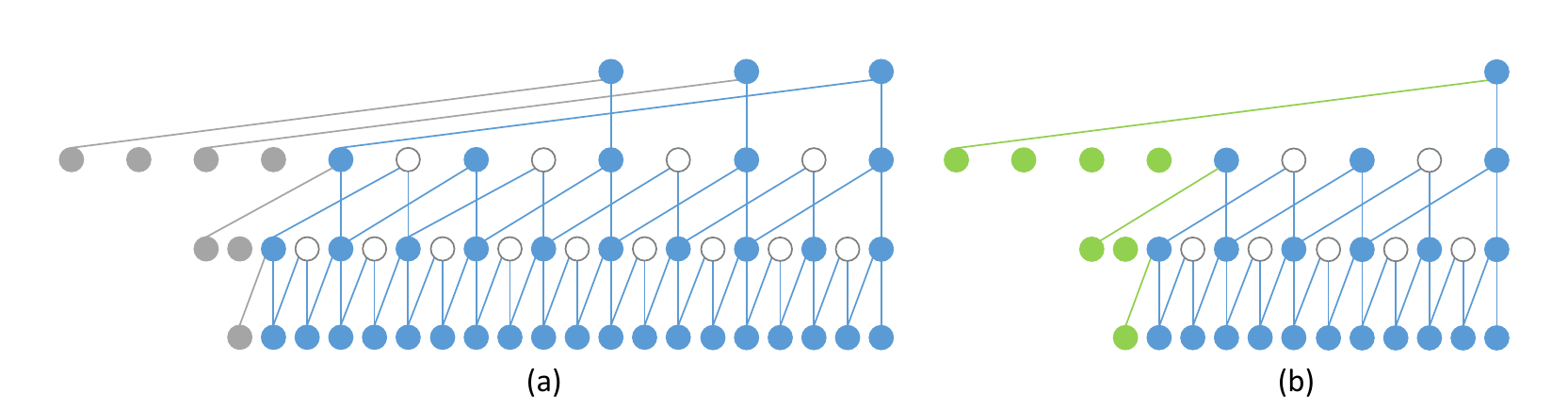}
  \caption{(a) Historical padding for training stage. (b) Historical context cache for inference stage.}
  \label{fig:train_and_inference}
\end{figure}

\begin{figure}[t]
  \centering
  \includegraphics[width=\linewidth]{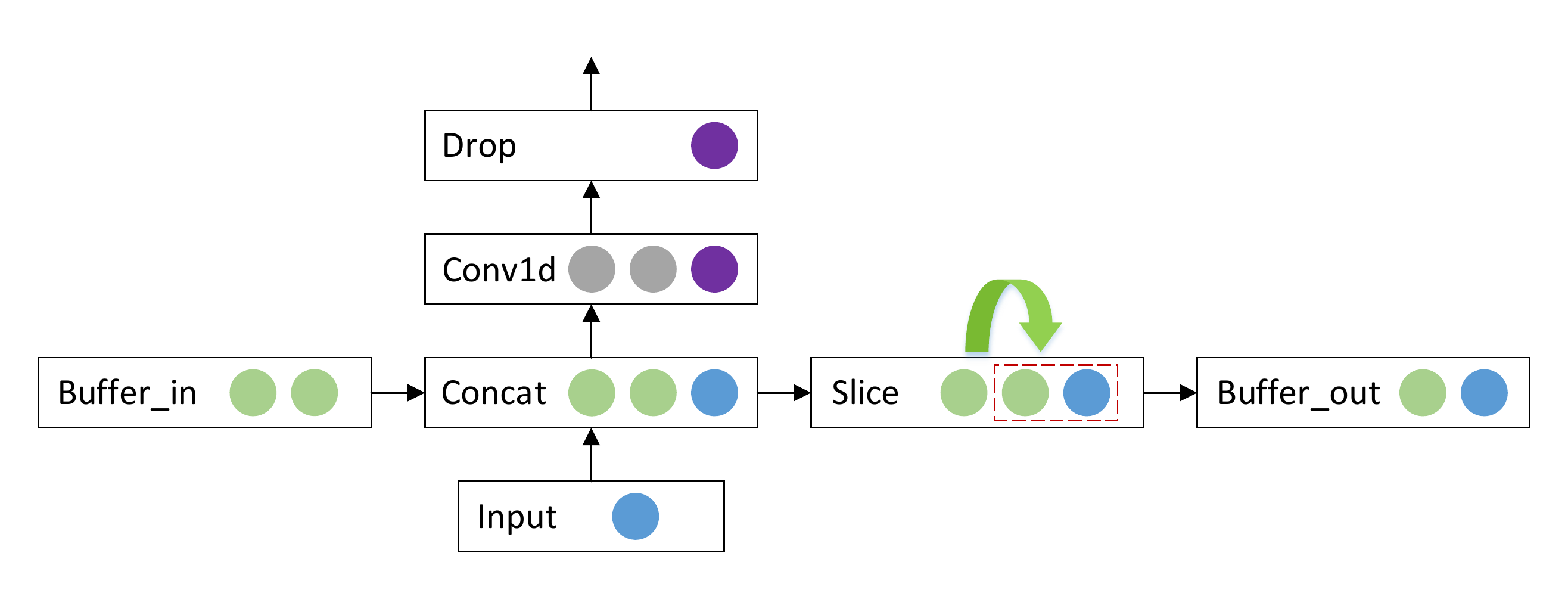}
  \caption{Context cache processing on inference time.}
  \label{fig:cache_incorp}
\end{figure}

For streaming inference we employed idea from ~\cite{transformerXL2019}. Paper represents how to extend the history context length which taken by a network into account if the network was being trained on fixed-size segments of data. The authors applied the mechanisms of hidden states reusing on the Transformer architecture and achieved considerable gain in the length of context remembered by the model. Intuitively it emulates the behavior of recurrent networks that save information between time steps. We tested this idea for TC Wave-U-Net in such a way that we do not try to extend, we try to keep the context size for inference as network used for training but for smaller chunk of data. It provokes the involved in processing the only part of the receptive field. Moreover, it leads to speed up of inference as calculations with floats should be dropped. In the order to keep the accuracy the same as for inference with whole shape we should "manually" reuse hidden states of each layer.
As mentioned before, the proposed method try to align conditions between training and inference modes. For training we use padding for each temporal convolution which could be calculated as \((kernel\_size-1)*dilation\). This situation is depicted on Figure~\ref{fig:train_and_inference}(a). The network goes through a fixed chunk of data passed by training framework, the size of which is bigger than the receptive field of the network and all layers are correctly trained. Hidden states here are passed in usual conditions from layer to layer automatically. For inference we are going to engage just a part of receptive field (Figure~\ref{fig:train_and_inference}(b)). If we pass cropped inputs with padding it will most likely lead to dramatic accuracy reduction, as weights did not learned to handle such situation. Instead of that we enable cache mechanism. For the first step we pass initiated by zeroes caches with the input. Cache size for each convolution layer is equal padding we used for training. It is concatenated with input and buffer for next iteration copied. Afterwards the whole tensor included input and history cache is convolved. Visual explanation of this sequence is presented on Figure~\ref{fig:cache_incorp}. At the end of the network we will have a stack of buffers with the length equal to the total number of convolution layers in encoder, bottleneck and decoder. Down and up blocks provoke shifting of starting point for the next layer where new cache history is captured.

As mentioned previously, the down-sampling mechanism is equal to the convolution with stride 2 and we have to take it into account when calculate the position for copy beginning. This leads to shape reduction.

\section{Experiments}

We conducted various experiments with our streaming TC Wave-U-Net model to challenge its speech enhancement quality and demonstrate efficiency of cache mechanism.

\subsection{Datasets}

The for training clean data we used audio speech datasets of Chineese language AISHELL-1, AISHELL-3, sets Librispeech (train-clean-360) and VCTK for the English language. The organizers were provided the lists permitted audio files from that datasets with loudness larger than 15dB that could be used for training procedure. The total duration of clean training speech was around 550 hours. The noise set was composed from two parts. Part 1 is selected from public noise datasets MUSAN and Audioset with total duration about 120 hours. The Part 2 was a real meeting room noise recordings. The total amount of provided clips was 98 items. Also we got simulated Room Impulse Responses (RIR) for the reverberation affect obtaining. Provided framework allowed to augment data on-the-fly during training. It considerably slow-down the training procedure. We generated and saved data locally.

\subsection{Evaluation Items}
Thus, the following denoising systems were taken to evaluate the efficacy of our proposed approach:

\begin{itemize}

  \item {\bf Baseline.} LSTM-based solution, that was provided by organizers itself. Basically the model process audio in time and frequency domains. The baseline has 8 channels raw audio input. It calculates inter-channel phase difference between pairs of microphone channels, complex STFT for input signal and pass he concatenated tensor through 3 LSTMs. Each recurrent layer has 512 hidden states and, finally, they followed by a projection layer. Resulted mask is applied on channel 1 of original signal and inverse STFT returns the output of baseline solution.
  \item {\bf Wave-U-Net.} The implementation which we took from open GitHub repository\footnote{\href{https://github.com/haoxiangsnr/Wave-U-Net-for-Speech-Enhancement}{https://github.com/haoxiangsnr/Wave-U-Net-for-Speech-Enhancement}}. It exploits 12 convolution layers in encoder and decoder. We just made changes in code to establish the ability of C-channel raw waveform processing instead of original mono channel input.
  \item {\bf TC Wave-U-Net.} Our proposed model with time domain input, casual convolutions in encoder decoder and bottle-neck and self-attentions in decoder.

\end{itemize}

\subsection{Experimental Setup}

We fed to network multi-channel audio recordings with additive random noise and convoluted with random RIRs. For training we capture random piece with 16384 samples from 16 kHz raw waveforms and used that data for input. Our network is constructed from 9 TC blocks in encoder and decoder, 1 casual convolution layer in bottleneck. Filter size for convolutions in encoder is 15 and 5 in decoder accordingly, as mentioned in paper ~\cite{sourceSeparation2018}. All channel sizes for encoder are 8, 24, 48, 72, 96, 120, 144, 168, 192, 216 (and 240 in bottleneck). For decoder they are repeated in reverse order. Dilation parameter for a each block is changed by the following way 1, 1, 1, 2, 4, 5, 16, 32, 64. The receptive field of the encoder is 1807 samples or ~112 ms.

\subsection{Learning Target}
For the learning objective we used weighted signal-to-distortion loss (\textsc{wSDR}) ~\cite{wsdr2019}. This is a time-domain loss function that could be defined by the following formula:
\begin{equation}
  L_{wSDR}(x,y,\overline{y}) :=  \alpha L_{SDR}(y,\overline{y}) + (1 - \alpha) L_{SDR}(z,\overline{z})
  \label{eq3}
\end{equation}
where \(L_{SDR}\) is a conventional signal-to-distortion (\textsc{SDR}) loss, \(x\) is a mixture signal is assumed as linear sum of clean speech signal \(y\) and original noise \(z\). Estimated noise \(\overline{z}\) given as \(\overline{z} = x - \overline{y}\). The \(\overline{y}\) is an enhanced audio. Taken this into account the energy ratio \(\alpha\) between dry speech \(y\) and noise \(z\) is defined as \(\alpha = \parallel{y}\parallel^2/(\parallel{y}\parallel^2 + \parallel{z}\parallel^2) \).

For the baseline system, we trained mean square error (\textsc{MSE}), signal-to-distortion  and \textsc{wSDR}. Our experiments show that \textsc{SDR} performed better that \textsc{MSE}  and \textsc{wSDR} was slightly better than \textsc{SDR}. In long training scenarios with \textsc{SDR} loss the model wasn't able to continue to converge while with \textsc{wSDR} loss even after 300 epoch the model steadily improved.
We train all our models using \(6\times\)Tesla V100. 

For the final training we choose \textsc{wSDR} loss and Adam optimizer with starting learning rate = \(10^{-3}\). The decayed learning scheduler was applied with minimum value \(10^{-8}\) after 250 epochs. Batch size is equal 1000.

The baseline was trained using default configuration parameters, \textsc{SDR} objective loss, Adam optimizer with learning rate = \(10^{-4}\) and with reduce on plateau scheduler. Chunks of 4 seconds was chooses by default as well. The dataset was identical to primary experiment. 

\subsection{Streaming Measurements Setup} \label{streaming}

We evaluated the performance of Baseline, Wave-U-Net and our TC Wave-U-Net in streaming mode. Wave-U-Net consist of simple convolution layers and does not satisfy the requirement of challenge for zero-look ahead. We changed code to conduct pseudo stream tensor processing. The chunk with 16384 samples ($\sim$1 sec) is fed to the model with shift 640 samples (40 ms) and 40 ms of predicted audio was taken from output. For the first times we provide chunk 40 ms, 80 ms, 120 ms etc. with zero-padded left part till it reached full size 16384.
For TC Wave-U-Net with cache we pass chunk window with 1024 samples (64 ms), and move the window on 640 samples (40 ms).

\section{Results}

\subsection{Signal Quality}

According to the conference rules, we push test dataset enhanced by our system to organizers that evaluate our resulting audio. Moreover, they provide mean opinion score (MOS) - measure of the human-judged overall quality of an audio. Each rater determines MOS, subjective speech MOS (S-MOS) and subjective noise MOS (N-MOS) for each cleaned recording. Next, confidence interval (CI) of MOS score is calculated. Each file is listened by more that 20 raters. In Table~\ref{tab:mos} we denoted the result of comparing of our enhanced audio with noised raw waveforms.

\subsection{Real Time Factor}
In Table~\ref{tab:rtf}, we report Real Time Factor (RTF) (processing time divided by audio duration) in relative scale where lower values indicate faster processing and lower user-perceived latency.

We can see that the network with cache enabled outperforms the models working with whole receptive field. It a bit slower than other networks but it compensated by better precise.
In the addition to mentioned above, in Table~\ref{tab:nonlinear_ma} we demonstrate the evaluation results of our model in streaming and non streaming modes. Baseline and vanilla Wave-U-Net with pseudo streaming (see~\ref{streaming}) are inferring close to the speed of TC Wave-U-Net with cache but with the lower quality than these models for non-streaming mode. PESQ for Baseline and Wave-U-Net with streaming 12\% and 8\% less, respectively, than non-streaming.  Also Wave-U-Net in streaming mode don’t meet the requirement of challenge for zero-look ahead.

Finally we converted TC Wave-U-Net to ONNX format to speed up the solution. It gave us {\bf 0.37} RTF on Intel Core i5 clocked at 2.4GHz.

\begin{table}[t!]
 	\caption{Results of evaluation test set enhanced with TC Wave-U-Net.}
	\label{tab:mos}
\centering
  \resizebox{0.7\columnwidth}{!}{
\begin{tabular}{lrrrr}
    \toprule
    \multicolumn{1}{l}{\bf Audio} &\multicolumn{4}{c}{\bf Metrics } \\ \midrule 
      & MOS & S-MOS & N-MOS & CI\\ 
    \midrule
    \textsc{Noisy}               & 2.56 & 2.93 & 3.03 & 0.02  \\
    \textsc{Enhanced}              & 2.90 & 3.05 & 3.05 & 0.04 \\
\bottomrule
\end{tabular}
}  
\end{table}

\begin{table}[t!]
 	\caption{Evaluation results of multichannel speech enhancement on Single Linear Nonuniform MA.}
	\label{tab:nonlinear_ma}
\centering
  \resizebox{1.0\columnwidth}{!}{
\begin{tabular}{lrrrrr}
    \toprule
    \multicolumn{1}{l}{\bf SetUp} &\multicolumn{1}{c}{\bf Audio} &\multicolumn{4}{c}{\bf Metrics } \\ \midrule 
    &   & PESQ & STOI & E-STOI & SI-SNR\\ 
    \midrule
    \textsc{Baseline}             & Noisy  & 1.515 & 0.823 & 0.690 & 4.474  \\
    \textsc{Baseline}             & Enhanced & 1.999 & 0.888 & 0.783 & 9.248 \\
    \textsc{Wave-U-Net}             & Enhanced & 2.132 & 0.888 & 0.797 & 9.528 \\
    \textsc{TC Wave-U-Net}             & Enhanced & 2.181 & 0.892 & 0.799 & 9.61 \\
    \textsc{TC Wave-U-Net cache enabled (Streaming)}             & Enhanced & \bf{2.192} & \bf{0.895} & \bf{0.802} & \bf{9.64} \\
\bottomrule
\end{tabular}
}  
\end{table}

\begin{table}[t!]
 	\caption{Real Time Factor for Streaming mode and Model sizes. Number of parameters (Millions).}
	\label{tab:rtf}
\centering
  \resizebox{1.0\columnwidth}{!}{
\begin{tabular}{lccc}
    \toprule
    \multicolumn{1}{l}{\bf Model} &\multicolumn{1}{c}{\bf RTF} &\multicolumn{1}{c}{\bf PESQ} &\multicolumn{1}{c}{\bf Size} \\ \midrule  
    \textsc{Baseline (Streaming)}             & \bf{0.7X}             & 1.76             & 8.68 \\
    \textsc{Wave-U-Net (Streaming)}           & 0.9X             & 1.93              & 10.14 \\
    \textsc{TC Wave-U-Net (Streaming)}         & 3.3X             & 2.18              & \bf8.31 \\
    \textsc{TC Wave-U-Net Cache enabled (Streaming)}   & X             & \bf{2.19}             & \bf8.31  \\
    \bottomrule
\end{tabular}
}  
\end{table}

\section{Conclusions}

This paper proposed our speech enhancement solution for Task1 on ConferencingChallenge2021. We provided the Wave-U-Net based network that outputs cleaned audio for passed raw waveform with reverberations and additive noise. Our system employed casual dilated convolutions for encoder, decoder and a bottleneck parts. It also involved self-attentions in decoder for better precise. We implemented historical cache and obtain fast streaming inference.

Our evaluation showed that adding cache mechanism for the model with large receptive field not only can reduce it for the expected one, but also reduce floating-point calculations, thereby improving the inference speed. Even more we have shown that compared to pure pseudo streaming, our proposed method provides the same quality as non-streaming model or does it a bit better. Due to the low time constraints, we aren't able to do the full research of the evaluation of calculation of the beginning of the cache, and have leaved it for the further research. Moreover, our final model is still training and we are waiting for final metrics.  

Future directions of this work could include experiments with deeper versions of this architecture, new versions of TC blocks, incorporation of Channel Attention and using complex ratio masking for signal enhancement.

\section{Acknowledgements}
The work of Artem Sokolov is partially supported by RSF (Russian Science Foundation) grant 20-71-10010.

\bibliographystyle{se/LaTeX/IEEEtran}

\bibliography{sebib}


\end{document}